\documentclass[a4paper, 12pt]{article}

\usepackage[latin1]{inputenc}

\usepackage[english, swedish]{babel}

\usepackage{graphicx}

\usepackage{amsmath}
\usepackage{latexsym}
\usepackage{amssymb}
\usepackage{epsfig}
\usepackage{cancel}

\numberwithin{equation}{section}
\numberwithin{table}{section}
\numberwithin{figure}{section}

\title{A Second Supersymmetry in Thermo Field Dynamics }

\author{Xerxes D. Arsiwalla \footnote{e-mail: xerxes@th.physik.uni-bonn.de } \\ \normalsize {\sl Physikalisches Institut der Universität Bonn,} \\
{\sl \normalsize Nußallee 12,
D-53115 Bonn,
Germany.} }

\date{}

\begin{document}

\selectlanguage{english}

\maketitle
\begin{abstract}
This article is an extension of the work done in \cite{partha}  by R. Parthasarathy and R. Sridhar. There they consider supersymmetry in an enlarged \\ thermal system (in a thermo field dynamic formulation) and show that this supersymmetry is not broken at finite temperature. Here we show, using an $SU(1,1)$ R-symmetry, that this system obeys a second supersymmetry. In addition, we proceed to see that this new supersymmetry also remains unbroken at  finite temperatures.  
\end{abstract}

\clearpage

\section{Constructing a second supersymmetry}
As in \cite{partha},  the components of the fermionic supercharges of the first \\ supersymmetry  are 
\begin{eqnarray}  
Q_+ = af^{\dagger} \hspace{0.15cm} ;\hspace{0.15cm} q_+ = \tilde{a}\tilde{f}^{\dagger} \hspace{0.15cm} ;  \hspace{0.15cm}  Q_- = a^{\dagger}f  \hspace{0.15cm} ; \hspace{0.15cm} q_- = \tilde{a}^{\dagger}\tilde{f}
\end{eqnarray} 
These are nilpotent operators acting on the enlarged Fock space $| n_B , \tilde{n}_B , n_F , \tilde{n}_F  \rangle$, and satisfy a $\mathbb{Z}_2$ - graded Lie algebra (that is, the structure $\{ O, O \} = E$, $[ O, E] = O$, $[ E, E] = E$ for even ( $E$ ) and odd ( $O$ ) operators ).   

Writing the above components in a vector form, we identify the \\ supercharge and its conjugate as  
\begin{eqnarray}  
 \left( \begin{array}{c} Q_+ \\ q_+ \end{array} \right) \quad \quad  and  \quad \quad  \left( \begin{array}{c} Q_- \\ q_- \end{array} \right)
\end{eqnarray}
Let us consider a $SU(1,1)$ transformation acting on our vector supercharge. This can be thought of as a R-symmetry transformation.  
\begin{eqnarray} 
\left( \begin{array}{cc} \alpha & \beta \\ \bar{\beta} & \bar{\alpha} \end{array} \right) \left( \begin{array}{c} Q_+ \\ q_+ \end{array} \right) = \left( \begin{array}{c} Q_1 \\ Q_2 \end{array} \right)  
\end{eqnarray}
( $\alpha , \beta \in \mathbb{C}$ and $| \alpha |^2 - | \beta |^2 = 1$ ) \quad  where 
\begin{eqnarray} 
Q_1 \equiv \alpha\hspace{0.15cm} Q_+ + \beta\hspace{0.15cm} q_+  \\
Q_2 \equiv \bar{\beta}\hspace{0.15cm} Q_+ + \bar{\alpha}\hspace{0.15cm} q_+ 
\end{eqnarray}
Similarly
\begin{eqnarray} 
\left( \begin{array}{cc} \alpha & \beta \\ \bar{\beta} & \bar{\alpha} \end{array} \right)^{\ast} \left( \begin{array}{c} Q_- \\ q_- \end{array} \right) = \left( \begin{array}{c} \bar{Q}_1 \\ \bar{Q}_2 \end{array} \right)  
\end{eqnarray}
 where 
\begin{eqnarray} 
\bar{Q}_1 \equiv \bar{\alpha}\hspace{0.15cm} Q_- + \bar{\beta}\hspace{0.15cm} q_-   \\  \bar{Q}_2 \equiv \beta\hspace{0.15cm} Q_- + \alpha\hspace{0.15cm} q_- 
\end{eqnarray}
Getting a new set of fermionic supercharge components  $Q_1 , Q_2, \bar{Q}_1 , \bar{Q}_2$; we now show that they indeed satisfy the extended superalgebra. In addition, the Hamiltonian is left invariant under the R-symmetry.  

Using the superalgebra relations of the first supersymmetry (as stated in \cite{partha} ), we see that the operators $Q_1 , Q_2, \bar{Q}_1 , \bar{Q}_2$ are indeed nilpotent; thereby converting bosons and fermions into each other when acting on the enlarged Fock space.  Proceeding to calculate the algebra generated by these operators, we obtain
\begin{eqnarray} 
\{ Q_1 , \bar{Q}_1 \} &=& | \alpha |^2 (N_B + N_F) + | \beta |^2 ( \tilde{N}_B + \tilde{N}_F )  \\
\{ Q_2 , \bar{Q}_2 \} &=& | \beta |^2 (N_B + N_F) + | \alpha |^2 ( \tilde{N}_B + \tilde{N}_F )  \\
\{ Q_1 , \bar{Q}_2 \} &=& \alpha \beta (N_B + N_F) + \alpha \beta  ( \tilde{N}_B + \tilde{N}_F )  \\
\{ Q_1 , Q_2 \} &=&  \{ \bar{Q}_1 , \bar{Q}_2 \} = 0  \\  
\left[ {\bf Q} , (N_B + N_F) \right] &=&  \left[ {\bf Q} , ( \tilde{N}_B + \tilde{N}_F ) \right] = 0 
\end{eqnarray}
where ${\bf Q}$ denotes any of $Q_1 , Q_2, \bar{Q}_1 , \bar{Q}_2$.  Hence, the operators $Q_1 , Q_2, \bar{Q}_1 , \bar{Q}_2 , \\ (N_B + N_F) , (\tilde{N}_B + \tilde{N}_F)$ too satisfy a $\mathbb{Z}_2$ - graded Lie algebra.   

Moreover, we can also see the anti-commutators between the two \\ generations of supercharges :
\begin{eqnarray} 
\{ Q_1 , Q_+ \} = \{ Q_1 , q_+ \} = \{ Q_2 , Q_+ \} = \{ Q_2 , q_+ \} = 0
\end{eqnarray}
and similarly their corresponding conjugate relations. These could be \\ interpreted as the absence of central charges. 

Note also the relations
\begin{eqnarray}
\{ Q_1 , Q_- \} = \alpha \hspace{0.15cm} (N_B + N_F)  \\
\{ Q_1 , q_- \} = \beta \hspace{0.15cm} ( \tilde{N}_B + \tilde{N}_F )  \\
\{ Q_2 , Q_- \} = \bar{\beta} \hspace{0.15cm} (N_B + N_F)  \\ 
\{ Q_2 , q_- \} = \bar{\alpha} \hspace{0.15cm} ( \tilde{N}_B + \tilde{N}_F ) 
\end{eqnarray}
and their corresponding conjugates. In the specific case of $\mathbb{Z}_2$ - graded Lorentz  algebras,  eqs. (1.15), (1.16), (1.17), (1.18)  would have been zero on the R.H.S. In other types of $\mathbb{Z}_2$ Lie gradings that need not be so. \\ However, at the moment, we do not have a definite interpretation of these R.H.S terms. 

Finally, putting together the full algebra, including both generations, we indeed have a super Lie algebra. It still remains to show, that the thermal system under consideration, obeys this second supersymmetry. 

In terms of the first supersymmetry, the total Hamiltonian is given by \cite{partha}
\begin{eqnarray}
\hat{H}_1 = \{ Q_+ , Q_- \} -  \{ q_+ , q_- \}
\end{eqnarray}
Under the R-transformation of the supercharges, the above Hamiltonian  \\ takes the form  
\begin{eqnarray}
\hat{H}_2 &=& \{ Q_1 , \bar{Q}_1 \}  -  \{ Q_2 , \bar{Q}_2  \}    \nonumber \\
&=&  \left( | \alpha |^2 - | \beta |^2 \right) \left[ (N_B + N_F) - ( \tilde{N}_B + \tilde{N}_F ) \right]    \nonumber \\
&=&  \hat{H}_1
\end{eqnarray}
thus remaining invariant under the R-symmetry transformation of the \\ supersymmetry  generators. Also, we have
\begin{eqnarray}
\left[ {\bf Q} , \hat{H}_2 \right] = 0
\end{eqnarray}
meaning that the supercharges of this second supersymmetry are all \\ conserved quantities. 

This concludes our argument that our thermal system confers to a second supersymmetry.

\section{Supersymmetry at finite temperatures}
In this section, we check how our second supersymmetry behaves at finite temperatures. The method followed is essentially the same as in \cite{partha}. 

First we look at the zero temperature case, where we get
\begin{eqnarray}
\langle 0 | \hat{H}_2 | 0 \rangle &=& 0  \\
and \hspace{2.5cm}  {\bf Q} | 0 \rangle &=& 0
\end{eqnarray}
implying that at zero temperature there is neither explicit nor spontaneous breakdown of the second supersymmetry. 

At finite temperatures, the thermal vacuum $| 0 (\beta ) \rangle$ has to be considered, which is annihilated by the Bogoliubov transformed annihilation operators. 

In \cite{partha}, three methods are considered. In the first method, the Fock space state vectors are taken to be temperature dependent, while the operators are not. Using this method for our second supersymmetry, we get similar results as in \cite{partha} :
\begin{eqnarray}
\langle 0 (\beta) | \hat{H}_2 | 0 (\beta) \rangle &=& 0  \\
and \hspace{2.5cm}  {\bf Q} | 0 (\beta) \rangle & \neq & 0
\end{eqnarray}
there is no explicit breaking, but only spontaneous breaking of the second supersymmetry. 

The second method involves a temperature dependence for both; the state vectors, as well as the operators acting on them. From our point of view, this method seems more appropriate. Now  the creation/annihilation operators have a $\beta$ - dependence given by the Bogoliubov transformations \cite{taka}. The canonical quantization relations still remain valid. Therefore, the  super Lie algebra too retains its form. The Hamiltonian can be seen to take the form
\begin{eqnarray}
\hat{H}_2 (\beta) =  \{ Q_1 (\beta), \bar{Q}_1 (\beta)\} -  \{ Q_2 , (\beta) \bar{Q}_2 (\beta) \}
\end{eqnarray}
The invariance of the Hamiltonian under the Bogoliubov transformation 
\begin{eqnarray}
\hat{H}_2 (\beta) = \hat{H}_2
\end{eqnarray}
can also be verified. Moreover, the Bogoliubov transformed supercharges $ {\bf Q} (\beta)  $ too are constants of motion. 

The following results are obtained 
\begin{eqnarray}
\langle 0 (\beta) | \hat{H}_2 (\beta) | 0 (\beta) \rangle &=& 0  \\
and \hspace{2.5cm}  {\bf Q} (\beta) | 0 (\beta) \rangle &=& 0
\end{eqnarray}
giving an explicitly as well as spontaneous unbroken second supersymmetry at finite temperatures. \cite{partha} and \cite{van} obtain similar results for the first \\ supersymmetry. 

The third method in \cite{partha} neglects the tilde operators, hence restricting to a sub-algebra of the super Lie algebra. Such a sub-algebra could still be chosen to contain one of the two supersymmetries. The $SU(1,1)$ R-symmetry is now reduced to a $U(1)$ phase. In a sense, this means, looking  at a sub-system of the full (mathematically) thermal system. Therefore, fixing to a sub-system comes at the price of losing some of the symmetry contained in the system as a whole. 

\section{Discussion and conclusions}
The central idea of this article is to show that the supersymmetric system given in \cite{partha} also contains a second supersymmetry. In order to do this, one has to find the appropriate R-symmetry, which maps the supercharges of one generation to that of another. At the same time, the R-action on the supercharges should leave the Hamiltonian invariant. We have shown that the $SU(1,1)$ R-symmetry does precisely that. 

Note that the supersymmetry considered here and in \cite{partha} is not the usual space-time  $\mathbb{Z}_2$ - graded supersymmetry of the Lorentz  algebra. The Lorentz superalgebra acts on Lorentz-covariant  quantum fields. In fact, the approach taken here, is far more modest. In our case, the superalgebra is actually a  $\mathbb{Z}_2$ - grading of the canonical commutator/anti-commutator algebra.  The Hamiltonian, is that of a free oscillator, with bosonic and fermionic degrees of freedom. The supercharges are defined at a given frequency $\omega$. Their action on the Fock space converts bosons and fermions ( at $\omega$ ) into each other, thus maintaining supersymmetric partner states at the same frequency. In a sense, this can be likened to the situation in supersymmetric quantum mechanics. 

In addition, for this second supersymmetry, we simply check its behaviour at finite temperatures, using the procedure in \cite{partha}. The corresponding \\ results are similar. At this stage, it is noteworthy to point out that at finite temperature, the supersymmetry constructed on Minkowski space quantum field theory \cite{buc}, \cite{buch}  shows interesting differences from the supersymmetry we have considered here.  There the supersymmetry is always broken \\ spontaneously in thermal states.

\section*{Acknowledgements}
We are grateful to Prof. R. Parthasarathy for proof reading the manuscript and for his encouraging remarks as well.  We would also like to thank Prof. R. Flume for informing us about the work of \cite{buc} and \cite{buch}.

\end{document}